 \documentclass[12pt]{article}

\usepackage[utf8]{inputenc}
\usepackage{graphicx}
\usepackage{amsmath}
\usepackage{amssymb}
\usepackage{color}

\usepackage[breaklinks]{hyperref}
\author{}
\title{}

\begin{document}

\author{Leonhard L\"ucken and Serhiy Yanchuk\\ 
\small{Institute of Mathematics, Humboldt University of Berlin,}\\
\small{Unter den Linden 6, 10099 Berlin, Germany}}

\title{Emergence of one- and two-cluster states in populations of globally
pulse-coupled oscillators}

\maketitle
%
%
%

\section{Introduction}

\begin{figure}
\includegraphics[width=0.9\textwidth]{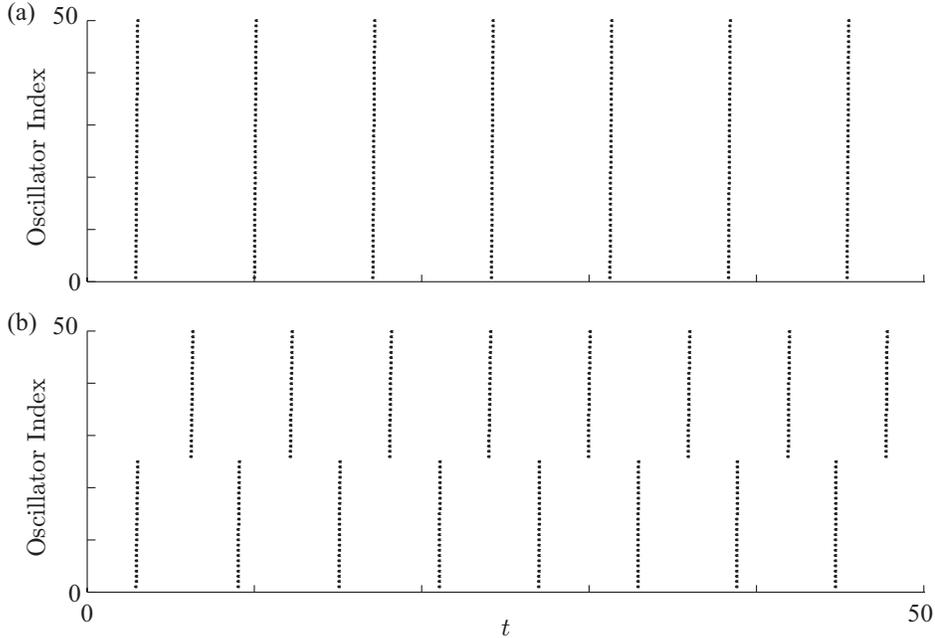} 
\caption{\label{figspikingdots} {Clusters in a population 
of 50 phase-oscillators. Dots indicate the times when an oscillator reaches
the threshold; (a) shows the firing pattern of a complete in-phase synchronized 
population (one-cluster), while (b) shows the firings in a symmetric two-cluster.}}
\end{figure}

Networks of coupled dynamical systems play an important role for all
branches of science \cite{Pikovsky2001,Strogatz2001,Strogatz2005,Tass1999}.
In the neuroscience, for instance, there is a need for modeling large
populations of coupled neurons in order to approach problems connected
with the synchronization of neural cells or other types of collective
behavior \cite{Timme2006,Popovych2006,Tass1999}. The investigation
of the dynamics of coupled lasers \cite{Wunsche2005,Fischer2006,Yanchuk2004,Yanchuk2006}
is important for many purposes including secure communication \cite{Cuomo1993,Kanter2008}
or high-power generation. The interacting biological, mechanical or
electrical oscillators \cite{Perlikowski2008,Perlikowski2010} belong
already to classical models for studying various aspects of collective
dynamics. In neural networks, 
the synchronous activity might be pathological \cite{Elble1990}, 
and hence, there was recently an increasing 
effort to control the desynchronization of populations of coupled oscillators.
In particular, the coordinated reset stimulation technique \cite{Tass1999,Tass2003}
proposes to establish a cluster-state in the network, in which 
the oscillator's phases split into several subgroups. 
This example illustrates the importance of the analysis of cluster formation
in coupled systems. 
Our paper investigates the connection between the properties of a single
oscillator, i.e. its sensitivity to stimulations,
and the formation of clusters in a globally coupled system of such oscillators. 
We show that by altering the shape of the sensitivity function, called 
the phase response function, different clusters in a network can be stabilized.
More presicely, we study a family of the phase response curves, which are
unimodal and turn to zero at the spiking moment. 
This choice is motivated by several well known neuron models. 
It appears that the position of the maximum of the unimodal sensitivity function 
with respect to the spiking point plays an important role for determining whether 
the system will synchronize or approach a two-cluster state (see Fig~\ref{figspikingdots}).
In particular, when the maximum 
of the sensitivity function is located in the second half of the 
period, the one-cluster (or completely synchronized) state acts as a global attractor.
In the case, when the sensitivity function reaches its 
maximum in the first half of the period, 
various two-cluster states become stable.

\subsection{Pulse-coupled oscillators}

In some coupled systems, e.g. neuron populations, the time, during
which the interaction effectively takes place is much smaller than
the characteristic period of oscillations. In such cases, it is reasonable
to approximate the interaction by an impact, i.e. by assuming that
the interaction is immediate. This approximation leads to models of
pulse-coupled oscillators, which have been widely used in the literature.
For example, Mirollo and Strogatz \cite{Mirollo1990} have shown,
that the complete synchronization (in this case it is equivalent to
the phase-locking) is stable and attracts almost all initial conditions
in the network of globally coupled Integrate-and-Fire (IF) oscillators
of the form 
\begin{equation}
\frac{dx_{j}}{dt}=S_{0}-\gamma x_{j},\quad x_{j}\in[0,1),\quad j=1,\dots,N\label{IF}
\end{equation}
with constants $S_{0}>\gamma>0$. 
One might refer to $S_0$ as input current and to $\gamma$ as the 
dissipation constant.
The following additional condition describes
the interaction: when $k$-th oscillator reaches the threshold
$x_{k}(t^{-})=1$, then positions of all remaining oscillators are
shifted accordingly to the rule 
\begin{equation}
x_{j}(t^{+})=
\min
\{ 
x_{j}(t)
+\varkappa,1
\},
\quad j\ne k
\label{xreset}
\end{equation}
with some small $\varkappa>0$ and the $k$-th oscillator resets
to $x_{k}(t^{+})=0$. It is shown in \cite{Mirollo1990}, that 
complete synchronization is achieved after a finite transient time.
The synchronization in a more general model of IF neurons has been
shown in \cite{Bottani1996}. Tsodyks \textit{et al.} have demonstrated
in \cite{Tsodyks1993} that the phase-locked state is unstable with
respect to inhomogeneity in the local frequencies, i.e. when the oscillators
become nonidentical.

A larger class of pulse-coupled models was studied in \cite{Goel2002,LaMar2010,Guardiola2000}.
In particular, Goel and Ermentrout \cite{Goel2002} obtained sufficient
conditions for the stability of a completely synchronous solution.
We introduce this class of models in the subsection \ref{PRC}.

The dynamics of pulse-coupled oscillators has been studied also
for systems with different topologies, i.e. ring 
topology \cite{Bressloff1997}, as well as for delayed
interactions \cite{Ernst1995}. 
Transient phenomena of randomly diluted
networks have been analyzed in \cite{Zumdieck2004}. Globally pulse-coupled
IF oscillators with a finite pulse-width have been considered in,
e.g. \cite{Zillmer2007,Abbott1993,Olmi2010}, where the interaction
pulse is assumed to have a shape $\frac{\alpha^{2}t}{N}e^{-\alpha t}$
with the width $\alpha$.

\subsection{Phase-response curve as a parameter \label{PRC}}

\begin{figure}[t]
\begin{centering}
\includegraphics[width=1\textwidth]{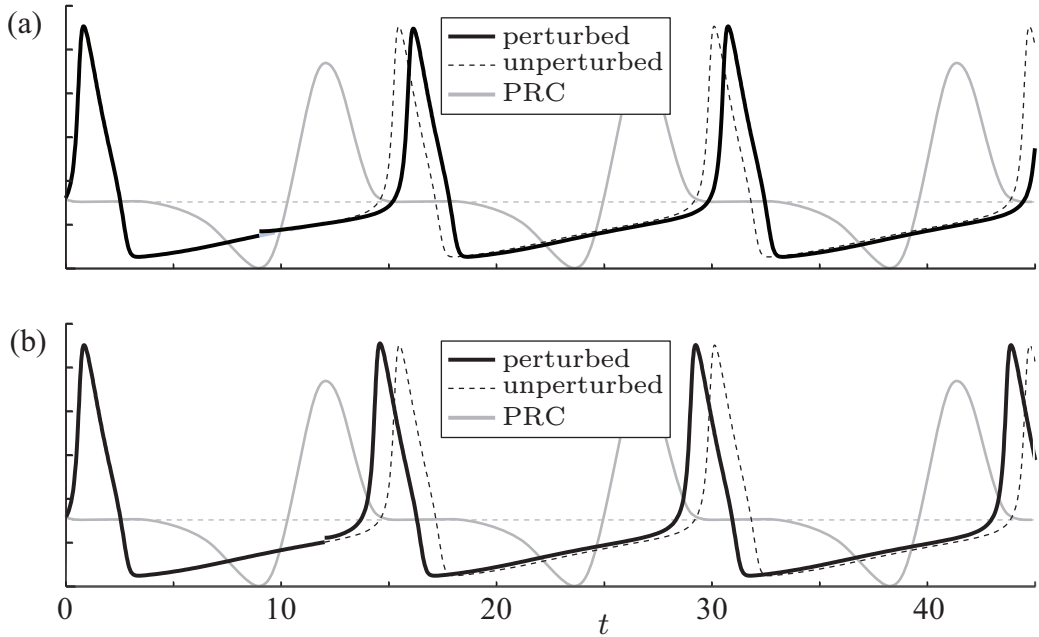} 
\par\end{centering}

\caption{ \label{figHHpuls} {Periodic spiking in a Hodgkin-Huxley neuron model. Solid black lines 
show the evolution of the voltage-component of the model when perturbed by a weak pulse at $t=9$ in (a), resp. $t=12$ in (b). 
The dashed black lines show the unperturbed oscillations. The PRC $Z(\varphi(t))$ of the unperturbed model is 
plotted solid grey and the dashed grey line corresponds to $Z=0$. 
The outcome of the perturbing pulse depends on the time $t$, or equivalently on the phase $\varphi(t),$ of its application. 
Either, the phase is delayed as in (a), i. e. $Z(\varphi(t))<0$, or it is forwarded as in (b), i. e. $Z(\varphi(t))>0$.}}

\end{figure}

In this subsection we introduce a general class of pulse-coupled phase
oscillators  \cite{Goel2002,Brown2004}. The
oscillator's motion between the spikes is described by the rule 
\begin{equation}
\frac{d\varphi_{j}}{dt}=\omega,\label{phieq}
\end{equation}
where $\varphi_{j}\in[0,2\pi]$. When 
$k$-th oscillator reaches the threshold at time $t$, i. e.  $\varphi_{k}(t^{-})=2\pi$,
it emits a spike to all other oscillators of the network, which 
are immediately resetted according to
\begin{equation}
\varphi_{k}(t^{+})=0;\quad\varphi_{j}(t^{+})=\varphi_{j}(t^{-})+\varkappa Z(\varphi_{j}(t^{-})),\quad j\ne k,\label{genphasereset}\end{equation}
 where $Z(\varphi)$ is called {\it phase response curve} (PRC). 
Effectively, this means that there is no coupling between two consecutive spiking events.
The coupling occurs only during the spike and acts through the resetting, since the
time of the resetting of the oscillator $j$ depends on the phase position of the oscillator $k$. 
The size of the phase-jump, that an oscillator performs, when stimulated by an 
incoming spike depends on its sensitivity to stimulation in its present state.
See figure \ref{figHHpuls} for an illustration.

\begin{figure}[t]
\begin{centering}
\includegraphics[width=1\textwidth]{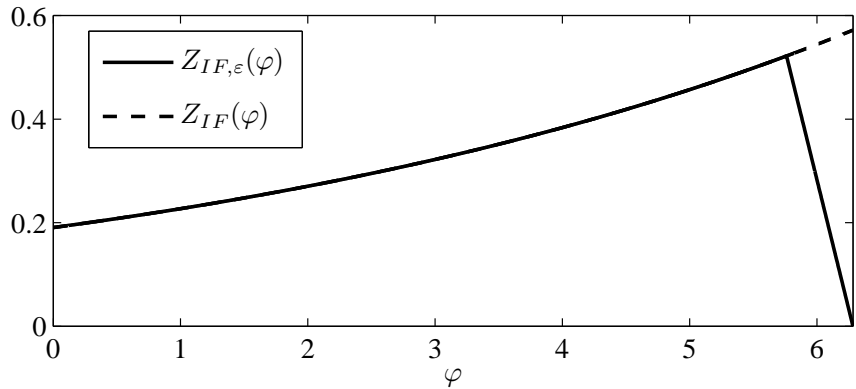} 
\par\end{centering}

\caption{ \label{figZIF} Phase-response curve for IF model (\ref{IF}). The
function $Z_{IF}(\varphi)$ measures the sensitivity of the system
to a small external perturbation at different positions $\varphi$.
The corrected function $Z_{IF,\varkappa}(\varphi)$ does not allow
the oscillators to be moved over the threshold by a spike. }

\end{figure}

Let us firstly show, that IF oscillators (\ref{IF})
can be  written in a form similar to 
(\ref{phieq})--(\ref{genphasereset}), see also \cite{Goel2002}.
For this, we rewrite (\ref{IF}) with respect to the phase coordinate
instead of the voltage coordinate. Indeed, the coordinate
$x_{j}$ in system (\ref{IF}) is supposed to describe the voltage
difference across the membrane of a neuron \cite{Izhikevich2005}.
The phase coordinate $\varphi_{j}$ should behave accordingly to (\ref{phieq})
with the frequency $\omega=2\pi/T$, where $T$ is the period of oscillations
without interaction and can be found from (\ref{IF}) \[
T=-\frac{1}{\gamma}\ln\left(1-\frac{\gamma}{S_{0}}\right).\]
 The corresponding transformation of variables $x=f(\varphi)$ can
be found from the condition \[
\frac{dx}{dt}=\frac{df}{d\varphi}\frac{d\varphi}{dt}=\frac{df}{d\varphi}\omega=S_{0}-\gamma f(\varphi),\]
 i.e. from the initial value problem \begin{equation}
\frac{df(\varphi)}{d\varphi}=\frac{T}{2\pi}(S_{0}-\gamma f(\varphi)),\quad f(0)=0.\end{equation}
 This gives the function \[
f(\varphi)=\frac{S_{0}}{\gamma}\left(1-\exp\left(-\frac{\gamma T}{2\pi}\varphi\right)\right),\]
which maps the interval $0\le\varphi\le2\pi$ into $0\le x\le1$.
In the transformed coordinates $\varphi_{j}$, the dynamics between
the spikes is described by (\ref{phieq}). It remains to specify the
dynamics at the threshold. Taking into account (\ref{xreset}), when
$k$-th oscillator reaches the threshold $\varphi_{k}(t^{-})=2\pi$ its
phase $\varphi_{k}$ resets to $\varphi_{k}(t^{+})=0$ and all other
oscillators have the impact \begin{eqnarray*}
\varphi_{j}(t^{+})=f^{-1}\left(x_{j}(t^{+})\right) & = & f^{-1}\left(x_{j}(t)+\varkappa\Delta(\varkappa,x_{j}(t))\right)\\
 & = & f^{-1}\left(f(\varphi_{j}(t))+\varkappa\Delta(\varkappa,f(\varphi_{j}(t)))\right),\end{eqnarray*}
 where $\Delta(\varkappa,x)=\min\{1,(1-x)/\varkappa\}\le1.$ In the
case of small $\varkappa$, i.e. the assumption of weak coupling holds,
the resetting rule can be approximated as \begin{equation}
\varphi_{j}(t^{+})=\varphi_{j}(t)+\varkappa\min\{Z_{IF}(\varphi_{j}(t)),(2\pi-\varphi_{j}(t))/\varkappa\},\label{phireset}\end{equation}
 where \begin{equation}
Z_{IF}(\varphi):=\frac{d(f^{-1})}{dx}(f(\varphi))=\frac{2\pi T}{S_{0}}\exp\left(\frac{T\gamma}{2\pi}\varphi\right).\label{zif}\end{equation}
Thus, with respect to the phase coordinates, the IF model (\ref{IF}) has the form (\ref{phieq}), (\ref{phireset}).
 In particular, the resetting rule is given by
the function \begin{equation}
Z_{IF,\varkappa}(\varphi)=\min\{Z_{IF}(\varphi_{j}(t)),(2\pi-\varphi_{j}(t))/\varkappa\},\label{prclif}\end{equation}
 which depends on the amplitude of the perturbation $\varkappa$.
Figure~\ref{figZIF} illustrates this function for $\varkappa=0.05$.
Practically, the PRC measures the sensitivity of the phase to external
perturbations.

We have shown above the specific example of pulse-coupled IF models 
and their reduction to pulse-coupled phase oscillators (\ref{phieq})--(\ref{genphasereset}). In fact,  
this procedure is also possible for higher-dimensional smooth systems, whenever
the oscillations correspond to a hyperbolic limit cycle, i.e.
in a generic case. More details can be found in  \cite{Goel2002,Brown2004,Hoppensteadt1997}. 
When the coupling is acting along one component, e.g. the voltage variable,
as often assumed in the case of neural populations, the PRC appears
as a scalar function of the phase. In the case of a higher-dimensional interaction, 
it should be considered more generally as a vector.

\begin{figure}
\begin{centering}
\includegraphics[width=1\textwidth]{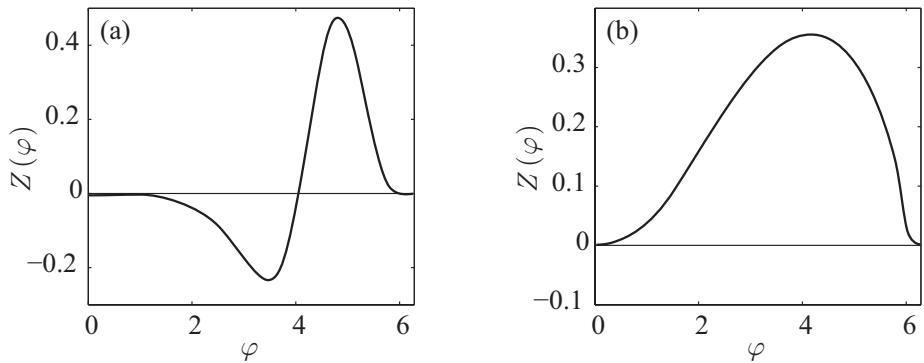} 
\par\end{centering}

\caption{ \label{figZex} Examples of different PRCs. (a) Hodgkin-Huxley model,
(b) Connor model. Note that the functions and their derivatives are
zero at the ends of the interval $\varphi=0$ and $\varphi=2\pi$
(adapted from \cite{Ermentrout1996}). }

\end{figure}

Examples of PRCs for different neuron models are shown in Fig.~\ref{figZex}.
Some more numerically and experimentally
obtained PRCs can be found in e.g. \cite{Goel2002,Brown2004,Ermentrout1996}.
The remarkable feature of many of such PRCs is that, contrary to the
IF model, their PRCs are independent on $\varkappa$ and admit zero
values at $\varphi=0$ and $\varphi=2\pi$. The conditions $Z(0)=Z(2\pi)=0$
are also reasonable from the neuroscientific point of view, since
they reflect the fact that the neurons are not sensitive to perturbations
during the spike (see Fig.~\ref{figZex}). Generally speaking, system
(\ref{phieq})--(\ref{genphasereset}) is a useful model, which possesses
quite a big generality by including the PRC as some ''infinite-dimensional''
parameter.

%


\subsection{System description}

\begin{figure}
\includegraphics[width=1\textwidth]{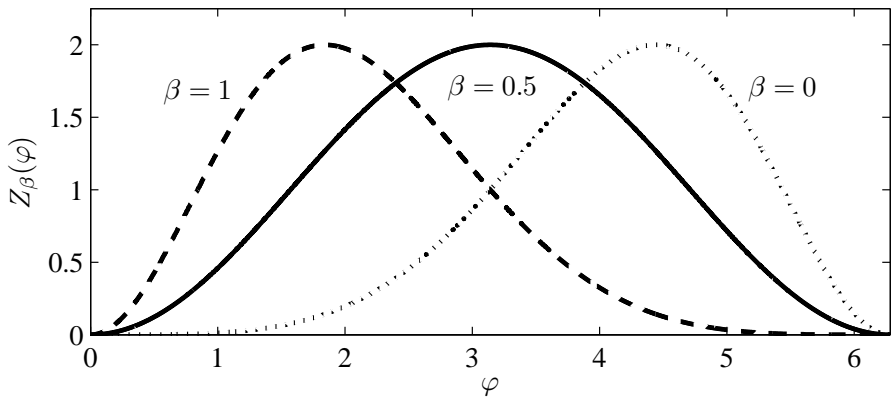} 
\caption{\label{figZbeta} Family of the unimodal PRCs $Z_{\beta}(\varphi)$,
see (\ref{Zbeta}).}

\end{figure}

Our main object of study is the following system of globally pulse-coupled
phase oscillators of the form \begin{equation}
\frac{d\varphi_{j}}{dt}=1\label{phieq1}\end{equation}
 with the resetting rule \begin{equation}
\varphi_{k}(t^{+})=0;\quad\varphi_{j}(t^{+})=\varphi_{j}(t^{-})+\frac{\varkappa}{N}Z(\varphi_{j}(t^{-})),\quad j\ne k,\label{genphasereset1}\end{equation}
where the velocity of the phase is assumed to be 1 without loss of
generality. We assume a fixed, positive overall coupling strength
$\varkappa>0.$ The impact is rescaled taking into account the number
of oscillators, see also \cite{Olmi2010}. 
In this study, we consider a one-parametric
family of the PRCs, which are positive and unimodal as shown in Fig.~\ref{figZbeta}.
The parameter $\beta\in[0,1]$ controls the position of the maximum,
namely, for larger $\beta$, the maximum is located in the domain
of small $\varphi$, which corresponds to a more sensitive excitatory
response of the system just after spike. For smaller $\beta$, the
system is more sensitive to perturbations shortly before the spike.
The value $\beta=0.5$ corresponds to an intermediate situation. We
assume also that $Z'(0)=Z'(2\pi)=0$, which is appropriate for a broad
class of experimental and analytically obtained PRCs (see Fig.~\ref{figZex}).

%

We note that the qualitative results reported in the paper are independent
on the exact expression for the PRC but rather on the shape of
the PRC and its behavior at $\varphi=0$ and $\varphi=2\pi$. 
Our particular choice is 
 \begin{equation}
Z_{\beta}(\varphi)=1-\cos\vartheta_{\beta}(\varphi),\quad\beta\in\left[0,1\right],\label{Zbeta}\end{equation}
 where \[
\vartheta_{\beta}(\varphi)=\left(1-\beta\right)\frac{\varphi^{2}}{2\pi}+\beta\left(2\pi-\frac{\left(\varphi-2\pi\right)^{2}}{2\pi}\right).\]
 In particular, $Z_{0.5}(\varphi)=1-\cos\varphi$.

System (\ref{phieq1})--(\ref{genphasereset1}) is equivalent
to an $(N-1)$-dimensional discrete dynamical system, which can be obtained
as a return map by considering its state each time when some of the
phases reaches a fixed value, e.g. $\varphi_{1}=2\pi$. Let us point
out how this map appears. Without loss of generality, we may assume
that the phases are ordered as \begin{equation}
2\pi=\varphi_{1}\ge\varphi_{2}\ge\cdots\ge\varphi_{N}\label{ordering}
\end{equation}
at $t=0$. 
 We will use the important property of (\ref{phieq1})--(\ref{genphasereset1})
that the oscillators do not overrun each other for all times if the
system size $N$ is sufficiently large. Indeed, since the inequality
\begin{equation}
\varphi_{j}+\frac{\varkappa}{N}Z(\varphi_{j})\ge\varphi_{j+1}+\frac{\varkappa}{N}Z(\varphi_{j+1})\label{orderpersistence}\end{equation}
 holds for sufficiently large $N$, the order of oscillators is preserved
during the spike. It is also evident, that the order is preserved
between the spikes as well. 
More exactly,  the inequality 
 $2\pi\ge\varphi_{1+l}\ge\varphi_{2+l}\ge\cdots\ge\varphi_{N+l}\ge0$
holds for all $t$,
where $l$ is some shift and the indices are considered modulo $N$.

Let us denote by $K_{1}$ the map, which maps the initial phases (\ref{ordering})
into the phases at the moment when the oscillator $\varphi_{2}$ reaches
the threshold, i.e. $\varphi_{2}=2\pi.$ It is easy to obtain that

\begin{align*}
 & K_{1}(\varphi_{1},\varphi_{2},\varphi_{3},\dots,\varphi_{N})=\\
 & =(2\pi-\mu(\varphi_{2}),2\pi,\mu(\varphi_{3})+2\pi-\mu(\varphi_{2}),\dots,\mu(\varphi_{N})+2\pi-\mu(\varphi_{2})),\end{align*}
where \begin{equation}
\mu(\varphi):=\varphi+\frac{\varkappa}{N}Z(\varphi).\label{eq:mu}\end{equation}
In a similar way, the mapping $K_{2}$ exists, which maps the phases
to the state, where the third oscillator is at the threshold and so
on. The composition of maps \begin{equation}
K=K_{N}\circ K_{N-1}\circ\cdots\circ K_{1}\label{K0}\end{equation}
gives the dynamical system on  $N$-dimensional torus $\mathbb{T}^{N}$
\begin{equation}
(\varphi_{1},\dots,\varphi_{N})\to K(\varphi_{1},\dots,\varphi_{N}),\label{K}\end{equation}
which 
maps the initial state  (\ref{ordering}) into a new state
after all $N$ oscillators have
crossed the threshold once and the first oscillator reaches again 
the threshold.
We call the map $K$  \textit{return map}.

In this paper, we will not use the explicit form of the mapping (\ref{K0}).
For our purposes it is important to conclude that the dynamics of
system (\ref{phieq1})--(\ref{genphasereset1}) are indeed equivalent
to some $(N-1)$-dimensional, discrete dynamical system on the $N$-dimensional
torus. The smoothness of this system depends on the smoothness of
its PRC function.


\section{Numerical results}


In order to detect the appearance of one- or two-cluster states, we
have numerically computed the order parameters \begin{equation}
R_{1}(t)=\left|\frac{1}{N}\sum_{k=1}^{N}e^{i\varphi_{k}(t)}\right|\label{op1}\end{equation}
 and \begin{equation}
R_{2}(t)=\left|\frac{1}{N}\sum_{k=1}^{N}e^{i2\varphi_{k}(t)}\right|.\label{op2}\end{equation}
 A perfect one-cluster state is characterized by $R_{1}=R_{2}=1$
and a perfect antiphase two-cluster is characterized by $R_{1}=0$
and $R_{2}=1$.
We present results of simulations for $\varkappa=0.5,$
but qualitatively we observe similar behavior for a broad range of
$\varkappa>0.$ 

\begin{figure}
\includegraphics[width=1\textwidth]{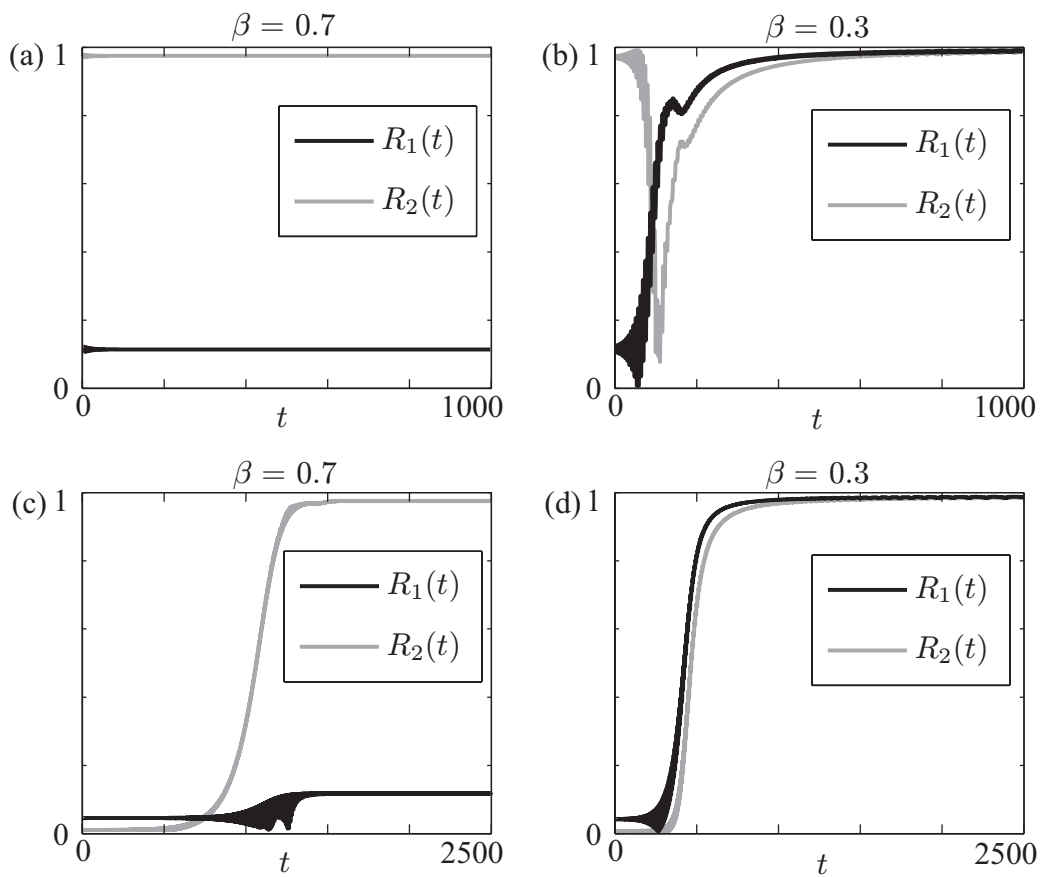} 
\caption{\label{OPtimelines} Behavior of the order parameters $R_{1}(t)$
and $R_{2}(t)$ for a trajectory starting in a vicinity of the two-cluster
state for (a) and (b). The lower panel (c) and (d) corresponds to
a trajectory starting in a vicinity of the splay state. Left figures
(a) and (c) correspond to the parameter value $\beta=0.7$, where
the two-cluster state is attracting and (b) and (d) to $\beta=0.3$,
where one-cluster state is attracting. }

\end{figure}

As shown in Fig.~\ref{OPtimelines}, we observe two qualitatively
different types of behavior depending on parameter $\beta$. For $\beta<0.5$,
i.e. when the maximum of the PRC is shifted to the right (see Fig.~\ref{figZbeta}),
the one-cluster state seems to be the attractor; for $\beta>0.5$ and the maximum
of the PRC is shifted to the left, a two-cluster state is attracting.
We have chosen initial conditions in a vicinity of a two-cluster state
in Fig.~\ref{OPtimelines}(a) and (b), therefore the initial values
of the order parameters are $R_{1}\approx0$ and $R_{2}\approx1$.
Figure \ref{OPtimelines}(b) shows how the instability of the two-cluster
state implies desynchronization transient, after which the system
is attracted to a synchronous one-cluster state. Similar behavior
occurs for other initial conditions. Figure~\ref{OPtimelines}(c)
and (d) illustrate the order parameters behavior for initial conditions
close to the splay state (a state, where the phases are distributed).
The initial values for the order parameters in the splay state are
close to zero, but after a transient, they approach again the same
asymptotic values as in (a) and (b).

\begin{figure}
\includegraphics[width=1\textwidth]{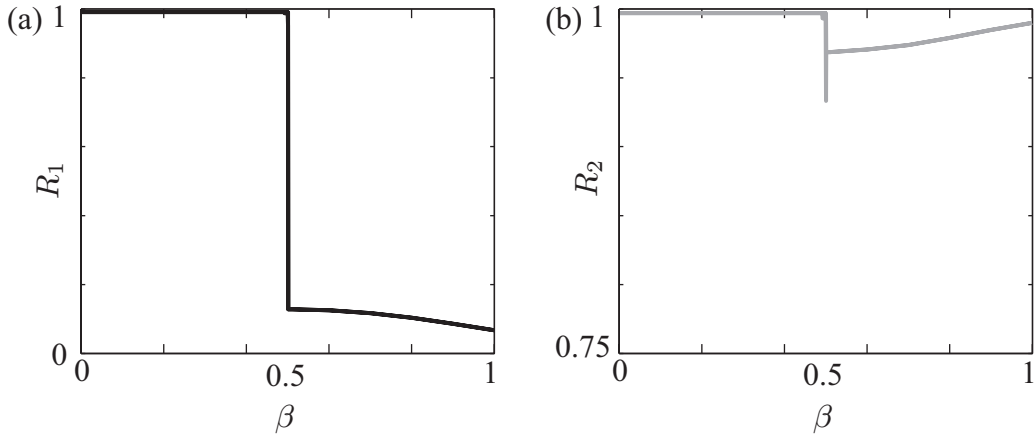} 
\caption{\label{OPvsbeta} Dependence of the asymptotic values for the order
parameter $R_{1}$ (a) and $R_{2}$ (b) on $\beta$. For the most
values of $\beta$, except $\beta=0.5$, the order parameters tend
to some constant value, when initialized near the splay state or the
symmetric two-cluster state. }

\end{figure}

A more complicated behavior occurs for the intermediate value of the
parameter $\beta=0.5$, i.e. when the PRC is symmetric. In this case,
the order parameters $R_{1}(t)$ and $R_{2}(t)$ do not approach some
asymptotic constant values but remain periodic in time. As a result,
the maximum asymptotic values of both $R_{1}$ and $R_{2}$ do not
coincide with the corresponding minimum values. This type of behavior
is observed for a very small parameter interval of order $10^{-3}$
around $\beta=0.5$. We discuss it  in Sec.~\ref{secbeta05} in more details.
Figure~\ref{OPvsbeta} summarizes the behavior of the order parameters
for different $\beta$.

%


\section{Appearance and stability properties of one-cluster state}


In an ideal one-cluster synchronized state, all oscillators have the
same phases $\varphi_{j}=\varphi_{s}$ for all $j$. This state is
a fixed point of the map (\ref{K}), because the PRC turns to zero
at $\varphi=2\pi$ and $\varphi=0$. This means that the
coupling vanishes for one-cluster state. More exactly, when an
oscillator $\varphi_{j}$ fires, i.e. $\varphi_{j}=2\pi$, all
other oscillators have the phase $2\pi$ and do not obtain the spike.
As a result, the period of this state is determined simply by the
uncoupled dynamics and equals $2\pi$.


\subsection{Inadequacy of the linear stability analysis}


In order to obtain conditions for the stability of one-cluster
state, one can examine the return map (\ref{K}). The linearization
of this return map around the one-cluster state gives then the corresponding
multipliers, which determine its local linear stability. As it is expected,
the local stability is governed by the  properties of the PRC
at $\varphi=0$ and $\varphi=2\pi$. This procedure has been done
in \cite{Goel2002}. Applying these results to our case, the resulting
conditions for the local linear stability of one-cluster state is
\begin{equation}
\left(1+\frac{\varkappa}{N}Z'(2\pi^{-})\right)^{l}\left(1+\frac{\varkappa}{N}Z'(0^{+})\right)^{N-l}<1,\quad l=1,N-1.\label{oneclusgoel}\end{equation}
 We observe that the necessary condition for the linear stability
is that the derivatives of $Z(\varphi)$ at the ends of the interval
$[0,2\pi]$ do not vanish. This is not the case for our PRC
 (\ref{Zbeta}). Hence, all associated multipliers
have modulus one and the linear stability analysis do not provide useful information
about the stability of one-cluster state.


\subsection{One-cluster state is a saddle point}


In this section we show that one-cluster state is a saddle point,
i.e. there are some arbitrary small perturbations of this state, which
grow with time. At the same time, some other small perturbations
decay.

\paragraph*{Existence of a local unstable direction. }

First of all, let us show that one-cluster state is unstable with
respect to the following special perturbation: \begin{equation}
\varphi_{1}=\varphi_{s}+\varepsilon,\quad\varphi_{2}=\cdots=\varphi_{N}=\varphi_{s}\label{pert1}\end{equation}
 with arbitrary small $\varepsilon>0$. During the period between
spikes, the dynamics is monotonous $\varphi_{j}(t)=\varphi_{s}+t$
for $j=2,\dots,N$ and $\varphi_{1}(t)=\varphi_{s}+\varepsilon+t$, thus,
 the distance between the phases remain constant. Without loss
of generality we may assume that \[
\varphi_{1}(0^{-})=2\pi,\quad\varphi_{2}(0^{-})=\cdots=\varphi_{N}(0^{-})=2\pi-\varepsilon.\]
 After the first oscillator moves over the threshold and resetting
occurs, the phases are as follows \[
\varphi_{1}(0^{+})=0,\quad\varphi_{2}(0^{+})=\cdots=\varphi_{N}(0^{+})=2\pi-\varepsilon+\frac{\varkappa}{N}Z(2\pi-\varepsilon)=\mu(2\pi-\varepsilon).\]
 The next resetting occurs at time $t_{1}=2\pi-\varphi_{2}(0^{+})=\varepsilon-\frac{\varkappa}{N}Z(2\pi-\varepsilon)$
when the group of $N-1$ synchronous oscillators reaches the threshold.
At this moment \[
\varphi_{1}(t_{1}^{-})=\varepsilon-\frac{\varkappa}{N}Z(2\pi-\varepsilon)>0,\quad\varphi_{2}(t_{1}^{-})=\cdots=\varphi_{N}(t_{1}^{-})=2\pi.\]
Now the group of $N-1$ synchronous oscillators is at the threshold.
The correct definition of the firing rule for this case can be naturally
obtained by extending it to the situation when all the oscillators
$\varphi_{2},\dots,\varphi_{N}$ in the cluster have slightly different
phases and then allowing the phases to converge to the same value.
This leads to the following resetting rule when passing the threshold
by the $N-1$ cluster: \begin{align}
 & \varphi_{1}(t_{1}^{+})=\mu^{N-1}\left(\varphi_{1}(t_{1}^{-})\right)=\mu^{N-1}\left(\varepsilon-\frac{\varkappa}{N}Z(2\pi-\varepsilon)\right),\label{resrule}\\
 & \varphi_{2}(t_{1}^{+})=\cdots=\varphi_{N}(t_{1}^{+})=0,\end{align}
 where $\mu^{N-1}$ denotes the superposition of $N-1$ functions
$\mu\circ\mu\circ\mu\circ\cdots\circ\mu$, where $\mu$ is defined
by (\ref{eq:mu}). The resetting
(\ref{resrule}) simply means that the function $\mu$ is applied
$N-1$ times (whenever an oscillator from the cluster $\varphi_{2},\dots,\varphi_{N}$
fires) in order to obtain the final position of $\varphi_{1}$.

\begin{figure}
\includegraphics[width=1\textwidth]{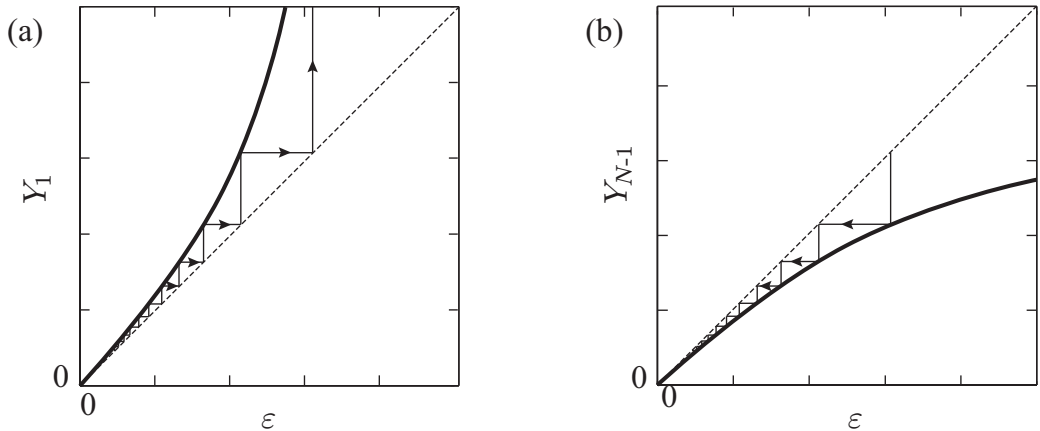} \caption{\label{cobwebY} Local Cobweb-Diagram of the functions $Y_{1}(\varepsilon)$
and $Y_{N-1}(\varepsilon)$ around $\varepsilon=0$. Iterations of
these maps determine the behavior of special perturbations to the
one-cluster state. (a): small perturbations grow with time; (b): small perturbations 
decay.}

\end{figure}

In this way, we obtain a mapping, which maps the initial size of the
perturbation $\varepsilon$ at time $t=0$ into its new size $Y_{1}(\varepsilon)$
at time $t_{1}$. The mapping is \begin{equation}
\varepsilon\to Y_{1}(\varepsilon)=\mu^{N-1}\left(\varepsilon-\frac{\varkappa}{N}Z(2\pi-\varepsilon)\right).\label{Y1}\end{equation}
 It is clear that $Y_{1}(0)=0,$ what corresponds to the invariance
of the one-cluster, and the stability properties of the origin of
(\ref{Y1}) determine the stability of the one-cluster state with
respect to the specific perturbation (\ref{pert1}) chosen. Up to
the linear level, the origin of (\ref{Y1}) is neutrally stable, i.e.
$Y_{1}'(0)=1$, which is clear, since the one-cluster state is linearly
neutrally stable. The second derivative of (\ref{Y1}) at $\varepsilon=0$
is nontrivial \[
Y_{1}''(0)=\varkappa Z''(0)-\frac{\varkappa}{N}\left(Z''(2\pi)+Z''(0)\right)\]
 and is positive for sufficiently large $N$ since $Z''(0)>0$ for
$\beta\in(0,1]$. Hence, for sufficiently large $N$, the origin of
(\ref{Y1}) is unstable, see Fig.~\ref{cobwebY}(a). \textit{This
leads to the local instability of one-cluster state  for all 
 $\beta\in(0,1]$.}
Accordingly to this, the distance of the advanced oscillator $\varphi_{1}$
from the remaining cluster will grow, but this growth is not exponential. 

\paragraph*{Existence of a local stable direction.}

Now let us show that the one-cluster state is locally stable with
respect to perturbations of the form \begin{equation}
\varphi_{1}=\varphi_{s}-\varepsilon,\quad\varphi_{2}=\cdots=\varphi_{N}=\varphi_{s}\label{per2}\end{equation}
 with $\varepsilon>0$. This can be shown similarly to the previous
case by obtaining the discrete mapping, which describes the dynamics
of the perturbation. In the case of perturbations (\ref{per2}), this
mapping reads \begin{equation}
\varepsilon\to Y_{N-1}(\varepsilon)=\mu\left(2\pi-\mu^{N-1}(2\pi-\varepsilon)\right)\label{Y2}\end{equation}
 and has the following properties\[
Y_{N-1}(0)=0,\]
\[
Y_{N-1}'(0)=1,\]
and\begin{equation}
Y_{N-1}''(0)=-\varkappa Z''(2\pi)+\frac{\varkappa}{N}\left(Z''(2\pi)+Z''(0)\right)\label{eq:D2-YN-1}\end{equation}
It implies that for sufficiently large $N$ the second derivative
is negative and the origin of the discrete mapping $\varepsilon\to Y_{N-1}(\varepsilon)$
is asymptotically stable (see Fig.~\ref{cobwebY}(b)). Hence, the
one-cluster state is stable with respect to perturbations of the form
(\ref{per2}). This, together with the instability with respect to
perturbations (\ref{pert1}), implies that the one-cluster state is
the saddle point in the phase space (see schematically Fig.~\ref{homoclinic}).

%

\paragraph*{Other stable and unstable local directions.}

In general, the two-cluster perturbations of the one-cluster state
are given by \begin{eqnarray}
\varphi_{1}\left(0\right) & = & ...=\varphi_{N_{1}}\left(0\right)=2\pi\label{2cl1}\\
\varphi_{N_{1}+1}\left(0\right) & = & ...=\varphi_{N}\left(0\right)=2\pi-\varepsilon,\label{2cl2}\end{eqnarray}
 where $N_{1}+N_{2}=N$. This means, there are $N_{1}$ oscillators
in the front-group and the remaining $N_{2}$ oscillators in the back-group.
The corresponding discrete 1-D systems, which describe the dynamics
of such perturbations are given by \[
\varepsilon\to Y_{N_{1}}(\varepsilon)\mbox{ and }\varepsilon\to Y_{N_{2}}(\varepsilon),\]
 where $Y_{j}(0)=0$, $\frac{d}{d\varepsilon}Y_{j}(0)=1$ for $j=1,...,N-1$
and \begin{equation}
\frac{d^{2}}{d\varepsilon^{2}}Y_{N_{1}}(0)=\frac{\varkappa}{N}\left(N_{2}Z''(0)-N_{1}Z''(2\pi)\right),\label{YNN1}\end{equation}
\begin{equation}
\frac{d^{2}}{d\varepsilon^{2}}Y_{N_{2}}(0)=\frac{\varkappa}{N}\left(N_{1}Z''(0)-N_{2}Z''(2\pi)\right).\label{YNN2}\end{equation}
The expressions (\ref{YNN1}) and (\ref{YNN2}) may have different
signs depending on the values of $N_{1}$, $N_{2}$, as well as the
second derivatives $Z''(0)$ and $Z''(2\pi)$. This implies the existence
of multiple unstable as well as stable directions to the one-cluster
solution, for more details, see section \ref{sec2cl}.


\subsection{Stable homoclinic orbit to one-cluster state}

\begin{figure}
\begin{centering}
\includegraphics[width=0.5\textwidth]{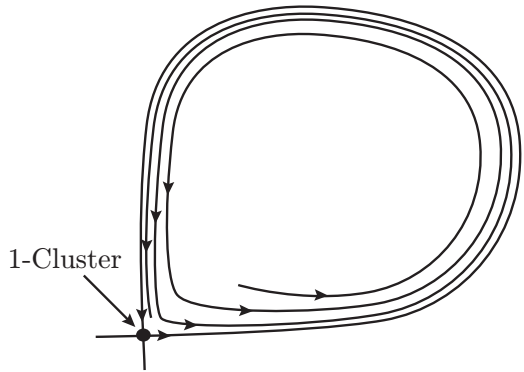} 
\par\end{centering}

\caption{\label{homoclinic} One-cluster state as a saddle point in the phase
space with a homoclinic loop. }

\end{figure}

Let us first note that the two-clusters of the form (\ref{2cl1})--(\ref{2cl2})
do not split with time. In geometric terms, this means, that the subspace
corresponding to such solutions 
is invariant. In particular, the subspace, which corresponds to $N_{1}=1$
and $N_{2}=N-1$ is invariant as well. Being restricted to this invariant
subspace, the one-cluster state is a saddle point, as we have shown
in the previous section. In Appendix~\ref{App} we prove that 
there exists a homoclinic orbit in this subspace, which connects
the both unstable and stable manifolds, see Fig.~\ref{homoclinic}.
In fact, as will be shown in Sec.~\ref{sec2cl}, the dynamics 
within the invariant subspace is given by the 1-D mapping 
shown in Fig.~\ref{2clustermap}(b).

%

\begin{figure}
\includegraphics[width=1\textwidth]{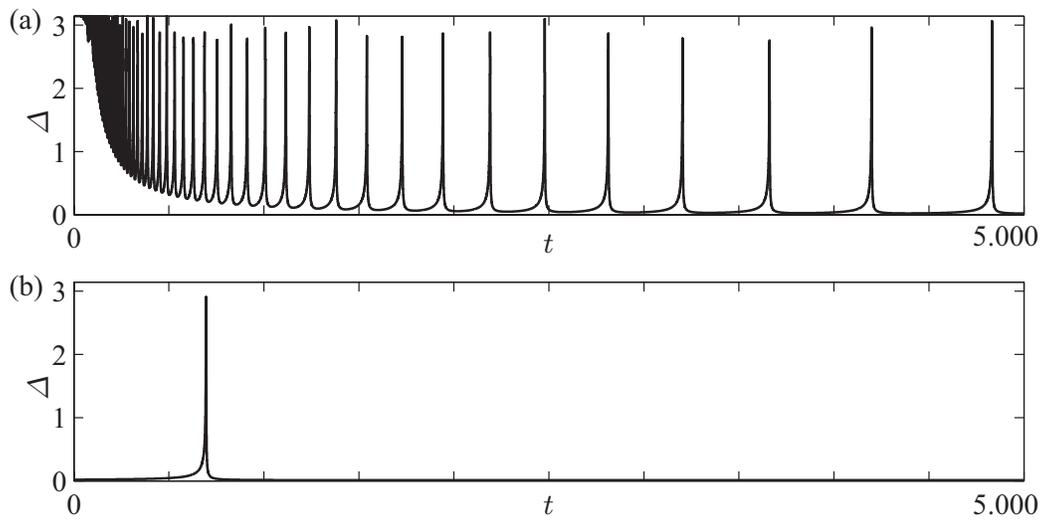} \caption{\label{spread} Width of the cluster $\Delta(t)=\max_{1\le i,j\le N}\left\{ \left|\varphi_{i}(t)-\varphi_{j}(t)\right|\right\} $
as a function of time. Figure (a) shows the behavior along the orbit
started at an initial condition close to the splay state (far from
the one-cluster). (b) shows the behavior along the orbit started close
to the state (\ref{pert1}). The behavior indicates the existence
of a stable homoclinic orbit. }

\end{figure}

Numerical calculations further supports this result and show that the invariant
set, which is composed of a homoclinic loop and the fixed point is
an attractor. Figure~\ref{spread} shows how the width of the cluster
changes as time evolves for some typical initial conditions. More
specifically, we compute \[
\Delta(t)=\max_{1\le i,j\le N}\left\{ \left|\varphi_{i}(t)-\varphi_{j}(t)\right|\right\} .\]
 One can clearly observe that the width tends eventually to zero interrupted
by some blowouts. The blowouts correspond to the events, during which
the first oscillator leaves behind the remaining cluster and makes
a rotation in the phase. After the rotation, it joins again the cluster
and becomes the ''last'' one. The time interval between such events
grows unboundedly with time supporting the homoclinic nature of the
attractor. Note that the width of the cluster should be nonzero in
order to observe this phenomenon, i.e. one should perturb the system
slightly from the fixed point, see Fig.~\ref{homoclinic}.

%

Finally, we would like to remark that the same methods allow proving
the existence of other homoclinic orbits, which correspond to two-cluster
perturbations (\ref{2cl1})--(\ref{2cl2}) with $N_{1}\ll N_{2}$.
Hence, one should rather speak about an attracting family of homoclinic
orbits.

\section{Two-cluster states}

\label{sec2cl}

Two-cluster state appears when the oscillators split into two groups 
(see Fig.~\ref{figspikingdots})
\begin{equation}
\varphi_{1}=\cdots=\varphi_{N_{1}}:=\psi_{1},\quad\varphi_{N_{1}+1}=\cdots=\varphi_{N_{1}+N_{2}}:=\psi_{2}.\label{2cl}\end{equation}

Contrary to one-cluster state, the two-cluster state must not be a
fixed point of the return map (\ref{K}). Indeed, when two clusters
appear, their relative behavior is then given by the following discrete
return map (by assuming that the return map is computed for $\psi_{1}=2\pi$
and $\psi_{2}<\psi_{1}$) \begin{equation}
\psi_{2}\to Y_{N_{1}}(\psi_{2}):=2\pi-\mu^{N_{2}}\left(2\pi-\mu^{N_{1}}(\psi_{2})\right).\label{2clmap}\end{equation}
This map has different properties depending on $N_{1}$, $N_{2}=N-N_{1}$
as well as on $\beta$. All such maps have zero fixed point corresponding
to the case when two clusters merge into one. One can obtain
\[
Y_{N_{1}}(0)=0,\quad Y_{N_{1}}(2\pi)=2\pi,
\]
\[
Y_{N_{1}}'(0)=1, \quad Y_{N_{1}}'(2\pi)=1,
\]
and 
\[
Y_{N_{1}}''(0)=\frac{\varkappa}{N}\left(N_{2}Z''(0)-N_{1}Z''(2\pi)\right),
\]
\[
Y_{N_{1}}''(2\pi)=\frac{\varkappa}{N}\left(N_{2}Z''(2\pi)-N_{1}Z''(0)\right).
\]

\begin{figure}
\includegraphics[width=1\textwidth]{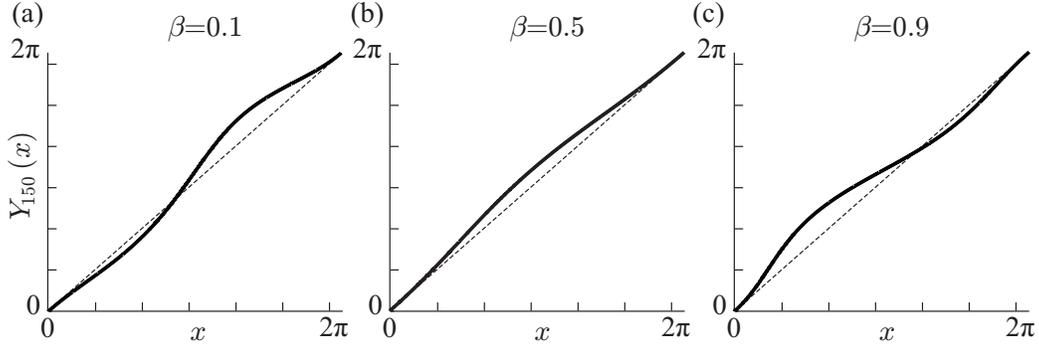} \caption{\label{2clustermap} Typical behavior of functions $Y_{N_{1}}(x)$
(see (\ref{2clmap})), which determine the behavior of two-clusters.
On the figure $N_{1}=150$ and $N=500$. }

\end{figure}

Figure~\ref{2clustermap} shows typical maps for three different
situations:\\
 (a) The map has an unstable fixed point inside the interval 
$[0,2\pi]$ and the
endpoints $x=0$ and $x=2\pi$ are asymptotically stable. Hence, within
the corresponding subspace, the one-cluster state is asymptotically
stable (similarly to Fig.~\ref{cobwebY}(b)). \\
 (b) The map has unstable fixed point at $x=0$ and stable at $x=2\pi$.
This case corresponds exactly to the case, when the one-cluster state
has a homoclinic orbit starting in $x=0$ and ending at $x=2\pi$
($0\sim2\pi$ on the torus).\\
 (c) The map has a stable fixed point inside the interval $[0,2\pi]$ 
and the endpoints
$x=0$ and $x=2\pi$ are unstable. Hence, within the corresponding
subspace, the one-cluster state is asymptotically unstable and the 
two-cluster stationary state is stable.

%

The fixed points of the map (\ref{2clmap}) give two-cluster stationary
states: \begin{equation}
\psi_{2}=Y_{N_{1}}(\psi_{2}).\label{2clmapfixp}\end{equation}

The condition for the merging of two cluster into one cluster is given
by the condition for the existence of the double root of the function
$Y_{N_{1}}(\psi)$ at $\psi=0$ or $\psi=2\pi$, i.e. $Y_{N_{1}}''(0)=0$
or $Y_{N_{1}}''(2\pi)=0$. This results into
\begin{equation}
\label{pitchfork}
N_{1}Z''(0)=N_{2}Z''(2\pi).
\end{equation}

\begin{figure}
\includegraphics[width=1\textwidth]{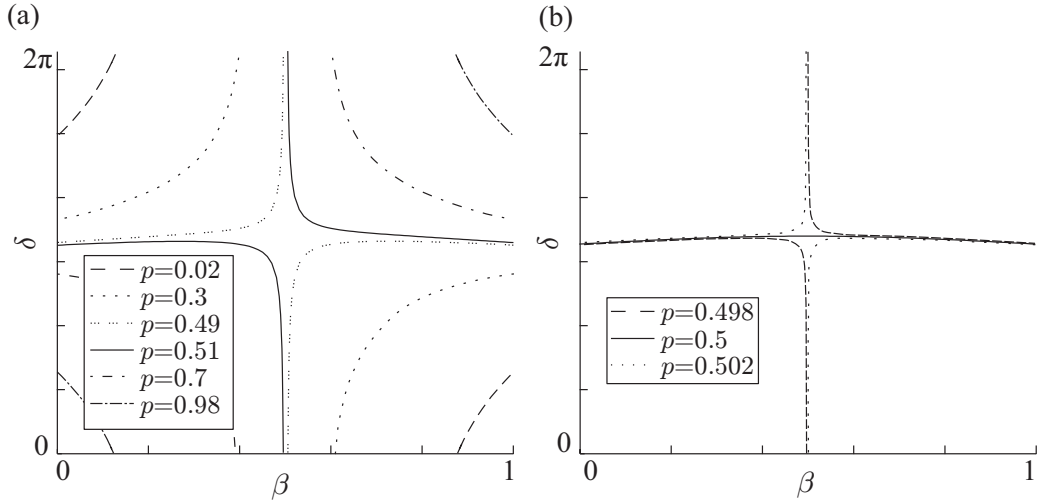}
\caption{\label{betavscl} Positions of the two-cluster states $\delta=2\pi-\psi_{2}$,
where $\psi_{2}$ are fixed points of (\ref{2clmapfixp}). Different
lines correspond to different cluster splittings, i.e. $N_{1}=pN$,
$N_{2}=(1-p)N$. At $\delta=0$ or $\delta=2\pi$, the corresponding
two-cluster is merging into the one-cluster. }

\end{figure}

 Expression (\ref{pitchfork}) determines also the moments when one-cluster
state undergoes pitchfork bifurcations. At such bifurcation, two different
nonsymmetric two-clusters bifurcate from the one-cluster state: one
with $N_{1}=pN$, $N_{2}=(1-p)N$, and another with $N_{1}=(1-p)N$,
$N_{2}=pN$. The bifurcation diagram in Fig.~\ref{betavscl} shows some of the
branches of two-clusters, which originate from $\psi_2=0$ or $\psi_2=2\pi$.

%

The pitchfork bifurcations for $\beta<0.5$ are subcritical. Namely,
the two-cluster states are unstable and they merge into the one-cluster
state. With increasing $\beta$ the one-cluster state becomes more
and more locally unstable by transforming stable directions into homoclinics
(see Fig.\ref{2clustermap}). In spite of this fact, we observe numerically,
that the invariant set, which is composed of the one-cluster state
and homoclinic connections is still attracting in the phase space.
All two-cluster states, which exist at this moment, are unstable.
As a result, one computes high values of the order parameters $R_{1}$
and $R_{2}$ on the numerically obtained figure \ref{OPvsbeta} for
$\beta<0.5$.


\subsection{Stability of two-cluster states}

\begin{figure}
\includegraphics[width=1\textwidth]{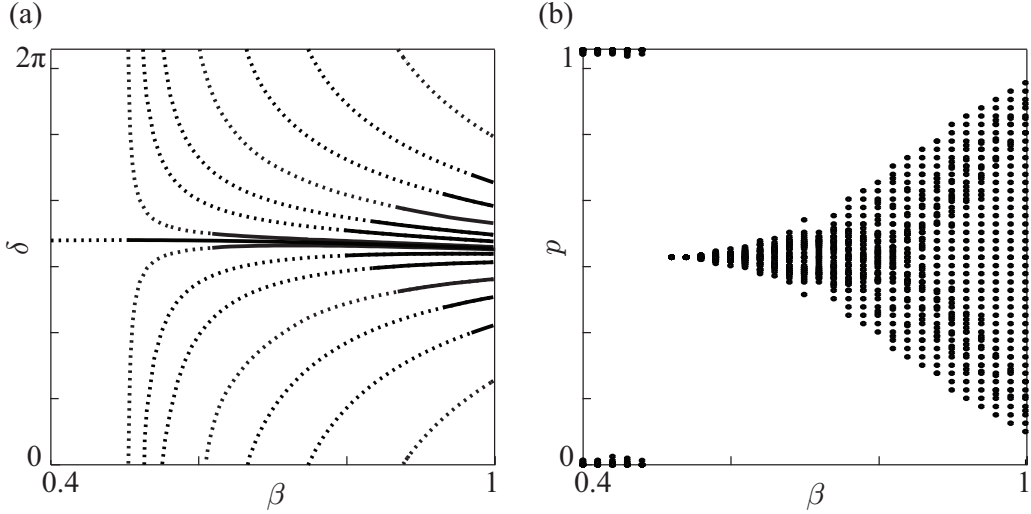} \caption{\label{last} Stability and existence of two-cluster states. (a) Solid
lines denote stable two-cluster stationary states and dashed - unstable.
The lines are shown only for selected values of $p=N_{1}/N$, while
the dense set of branches for all possible $p$ exist. Figure (b)
shows which two-clusters are stable in dependence on $\beta$ (obtained
numerically). $p=0.5$ corresponds to the symmetric cluster and $p\ne0.5$
to nonsymmetric clusters. }

\end{figure}

For $\beta>0.5$, the invariant set composed of one-cluster state
and homoclinic orbits losses its stability and two-cluster states
emerge, which are asymptotically stable. Numerical results in Fig.~\ref{last}
show which two-clusters are stable depending on the parameter $\beta$.
In general, for $\beta$ closer to 0.5, the symmetric clusters with
$p\approx0.5$ are stable. As $\beta$ increases, the more asymmetric
clusters stabilize as well. This implies that the PRCs with the maximum,
which is shifted to the left favor the coexistence of a large number
of stable branches of two-clusters.

%


\section{Intermediate state for symmetric PRC with $\beta=0.5$ \label{secbeta05}}

\begin{figure}
\includegraphics[width=1\textwidth]{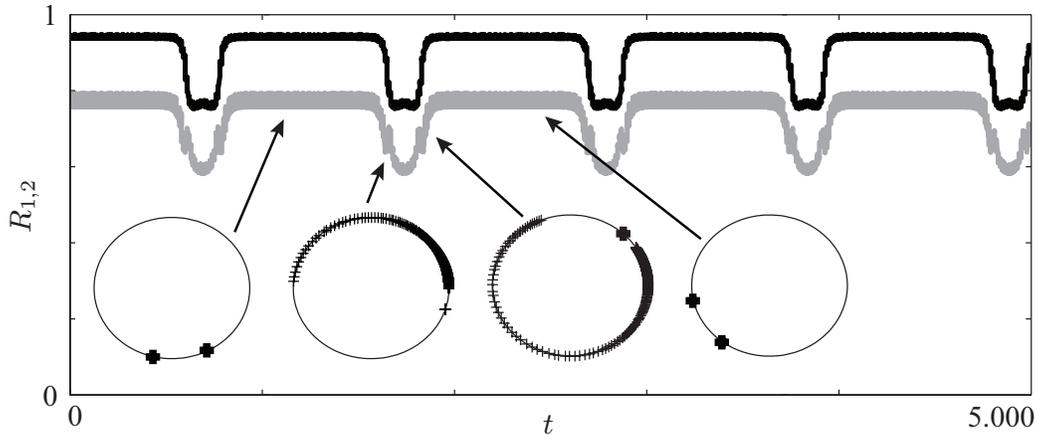} \caption{\label{intermediate} Nonstationary behavior of the order parameters
$R_{1}$ and $R_{2}$ with time for $\beta=0.5001$. One observes
periodic restructuring of two-clusters. }

\end{figure}

The case of symmetric PRC for $\beta=0.5$ is degenerate. When increasing
$\beta$ through $0.5$, the homoclinic sets including the one-cluster
state become unstable and a two-cluster state becomes stable as it
is described in the previous section. The numerical calculations for
$\beta=0.5$ show nonstationary dependence of the order parameters
on time, see Fig.~\ref{intermediate}. One observes periods of time,
when two-clusters persist. These periods are characterized by almost
constant order parameters. The periodic blowouts of the order parameters
correspond to the behavior, during which the oscillators from the
advancing cluster spread over a big part of the phase circle and finally
form another cluster behind (see the inset in Fig.~\ref{intermediate}).

%


\section{Conclusions}


In this paper we have studied the asymptotic behavior of a system
of globally pulse-coupled phase oscillators (\ref{phieq1})--(\ref{genphasereset1})
with the phase response function, which is positive, unimodal, and
turns zero at the threshold together with its first derivative. In
particular, we considered the question how the position of the maximum
of the PRC influences the dynamics of the coupled system.

We have numerically observed, that for the PRCs with the maximum shifted
to the right (for our model, it corresponds to $\beta<0.5$), a one-cluster
state becomes apparently stable. More detailed analysis reveals that
the one-cluster state is, in fact, asymptotically locally unstable,
i.e. a generic small perturbation will grow with time. Moreover, we
show that trajectories of the system has a behavior, which is characterized
by long-time intervals when the system stays close to the one-cluster
state and long excursions away from the one-cluster state (see Fig.~\ref{spread}).
The excursions become less and less frequent with time. This behavior
is explained by the existence of the family of homoclinic orbits to
the one-cluster state, which altogether form an attracting set in
the phase space of the system.

In the case, when the maximum of the PRC is shifted to the left, i.e.
the oscillators are mostly sensitive to perturbations in the phase
just after the threshold, the one-cluster state no more dominates
the dynamics and various stationary two-cluster states become stable.
These two-cluster states appear in pitchfork bifurcations from the
one-cluster state as parameter $\beta$ increases. 
First, at $\beta=0.5$, there appears a symmetric two-cluster
with equal number of oscillators in each cluster. With further increasing
$\beta$ more and
more asymmetric clusters appear and become stable leading to the increasing
coexistence of stable two-clusters. 

\section{Appendix: Existence of a homoclinic orbit \label{App}}

\textbf{Theorem.} \textit{For $\beta\in\left(0,1\right)$ there exists
 $N_{0},$ such that
for populations of size $N>N_{0},$
system (\ref{K}) possesses a homoclinic trajectory, which connects
the one-cluster stationary state. The homoclinic trajectory has the
form 
\begin{equation}
2\pi=\varphi_{2}(n)=\cdots\varphi_{N}(n)\ne\varphi_{1}(n),
\label{homorb}
\end{equation}
where $\lim_{n\to-\infty}\varphi_{1}(n)=0^+$ and $\lim_{n\to+\infty}\varphi_{1}(n)=2\pi^-$.}

  {\it Proof.}
Fix $\beta\in\left(0,1\right).$ We will consider 
\[
Y_{1}\left(N,x\right)  =  2\pi-\mu\left(N,2\pi-\mu^{N-1}\left(N,x\right)\right),
\]
where $\mu^{j}\left(N,x\right)$ denotes the $j$-th iteration of
\[
x\mapsto\mu\left(N,x\right)=x+\frac{\varkappa}{N}Z_{\beta}\left(x\right).\]
The map $Y_{1}\left(N,x\right)$ describes the evolution of the distance
$x\in\left(\text{0,2\ensuremath{\pi}}\right)$ during a time interval
in which all oscillators of a population $
\varphi_{1}=x;\,\,\,\,\varphi_{2}=...=\varphi_{N}=2\pi,$
emit exactly one spike. Homoclinicity then is equivalent to
 \[
Y_{1}^{k}\left(N,x\right)\to2\pi\mbox{, for all }x\in\left(0,2\pi\right)\mbox{, as }k\to\infty,
\]
where $Y_{1}^{k}\left(N,x\right)$ denotes the $k-th$ iteration of
$x\mapsto Y_{1}\left(N,x\right).$ Analogously to the analysis of
section \ref{sec2cl}, we find that 
\begin{eqnarray*}
 & &Y_{1}\left(N,0\right)=0;\quad Y_{1}\left(N,2\pi\right)=2\pi,\\
 & & Y_{1}^{\prime}\left(N,0\right)=Y_{1}^{\prime}\left(N,2\pi\right)=1,\\
 & &Y_{1}^{\prime\prime}\left(N,0\right)>0,\quad Y_{1}^{\prime\prime}\left(N,2\pi\right)>0.\end{eqnarray*}
Here and in the following, primes denote the derivatives with respect to the
second argument (phase). 
For fixed $N,$ there exists a rejecting region $\left(0,\varepsilon_{N}\right)$
where $Y_{1}^{\prime\prime}\left(N,x\right)>0,$ for $x\in\left(0,\varepsilon_{N}\right)$
and an attracting region $\left(2\pi-\varepsilon_{N},2\pi\right)$
with $Y_{1}^{\prime\prime}\left(N,x\right)>0,$ for $x\in\left(2\pi-\varepsilon_{N},2\pi\right).$
This gives:\[
Y_{1}^{k_{0}}\left(N,x\right)>\varepsilon_{N},\]
for $x\in\left(0,\varepsilon_{N}\right)$ and some finite $k_{0}=k_{0}\left(N,x\right)\in\mathbb{N},$
and\[
Y_{1}^{k}\left(N,x\right)\to2\pi,\]
for $k\to\infty$ and $x\in\left(2\pi-\varepsilon_{N},2\pi\right).$
Our goal is to show, that there exists a uniform $\varepsilon_{0}>0,$
such that for all $N>N_{0}:
$ \begin{eqnarray*}
&&Y_{1}^{\prime\prime}\left(N,x\right)  >  0\quad \mbox{ for }x\in\left(0,\varepsilon_{0}\right)\quad \mbox{ and}\\
&&Y_{1}^{\prime\prime}\left(N,x\right)  >  0\quad \mbox{ for }x\in\left(\varepsilon_{0},2\pi-\varepsilon_{0}\right),
\end{eqnarray*}
and such that for all $N>N_{0}$ and all $x\in\left[\varepsilon_{0},2\pi-\varepsilon_{0}\right]:$
\[
Y_{1}\left(N,x\right)>x+\Delta_{N},\]
with \[
\Delta_{N}:=\min_{x\in\left[\varepsilon_{N},2\pi-\varepsilon_{N}\right]}Y_{1}\left(N,x\right)-x>0.\]
Thus, any $x\in\left(0,2\pi\right)$ will reach the attracting region
$\left(2\pi-\varepsilon_{0},2\pi\right)$ within a finite number of
iterations of $x\mapsto Y_{1}\left(N,x\right).$ Let us write \[
Y_{1}\left(N,x\right)=\tilde{Y}_{1}\left(x\right)+\frac{1}{N}w(N,x),\]
where $\tilde{Y}_{1}\left(x\right)  =  x+\varkappa Z_{\beta}\left(x\right)$
is independent of $N.$ For $\tilde{Y}_{1}$ we have
\begin{eqnarray*}
& & \tilde{Y}_{1}\left(x\right)>x\quad \mbox{ for }x\in\left(0,2\pi\right),\\
 && \tilde{Y}_{1}\left(0\right)=0,\,\,\tilde{Y}_{1}\left(2\pi\right)=2\pi,\\
 & &\tilde{Y}_{1}^{\prime}\left(0\right)=\tilde{Y}_{1}^{\prime}\left(2\pi\right)=1,\end{eqnarray*}
This implies for\[
w\left(N,x\right)=N\left(Y_{1}\left(N,x\right)-\tilde{Y}_{1}\left(x\right)\right),\]
that 
\begin{eqnarray*}
&&w\left(N,0\right)=w\left(N,2\pi\right)  =  0,\\
&&w^{\prime}\left(N,0\right)=w^{\prime}\left(N,2\pi\right)  =  0.
\end{eqnarray*}
We will show, that the region $\left[0,\varepsilon_{N}\right]$ may
be chosen as $\left[0,\varepsilon_{0}\right],$ independently on large
$N.$ The analysis for the other region $\left[2\pi-\varepsilon_{0},2\pi\right]$
can be done similarly. Around $x=0,$ we have the following
representation of $Y_{1}\left(N,x\right)$:
\[
Y_{1}\left(N,x\right)  =  Y_{1}\left(N,0\right)+Y_{1}^{\prime}\left(N,0\right)x+\frac{x^{2}}{2}Y_{1}^{\prime\prime}\left(N,\xi_{N}\right)
\]
\[  =  \tilde{Y}_{1}\left(0\right)+\tilde{Y}_{1}^{\prime}\left(0\right)x+\frac{x^{2}}{2}\tilde{Y}_{1}^{\prime\prime}\left(\xi_{N}\right)+\frac{1}{N}\left(w\left(N,0\right)+w^{\prime}\left(N,0\right)x+\frac{x^{2}}{2}w^{\prime\prime}\left(N,\xi_{N}\right)\right)
\]
\[
  =  x+\frac{x^{2}}{2}\left(\tilde{Y}_{1}^{\prime\prime}\left(\xi_{N}\right)+\frac{1}{N}w^{\prime\prime}\left(N,\xi_{N}\right)\right)
\]
for some $\xi_{N}\in\left[0,\varepsilon\right].$ 
Further it holds $\tilde{Y}_{1}^{\prime\prime}\left(0\right)>0.$
This means, there exists an $\varepsilon_{0}>0,$ such that for $x\in\left[0,\varepsilon_{0}\right],$
$\tilde{Y}_{1}^{\prime\prime}\left(x\right)>0.$ Now we construct
an $N$-independent lower bound for $w^{\prime\prime}\left(N,x\right)$
in $x\in\left[0,\varepsilon_{0}\right],$ where $\varepsilon_{0}$
will be further altered in the analysis without always choosing a
new notation. In other words, we claim that there exists $c_{0}\in\mathbb{R}$
with 
\begin{equation}
\liminf_{N\to\infty}\left(\min_{x\in\left[0,\varepsilon_{0}\right]}w^{\prime\prime}\left(N,x\right)\right)  >  c_{0}.\label{eq:lower-bound-for-D2w}
\end{equation}
We have
\[
w\left(N,\varepsilon\right)  =  N\left(Y_{1}\left(N,x\right)-\tilde{Y}_{1}\left(x\right)\right)
\]
\[ =  N\left(2\pi-\mu\left(2\pi-\mu^{N-1}\left(N,x\right)\right)-x-\varkappa Z_{\beta}\left(x\right)\right)
\]
 \[
 =  N\left(\mu^{N-1}\left(N,x\right)-\frac{\varkappa}{N}Z_{\beta}\left(2\pi-\mu^{N-1}\left(N,x\right)\right)-x-\varkappa Z_{\beta}\left(x\right)\right)
\]
 \[ =  N\left(\frac{\varkappa}{N}\sum_{j=0}^{N-2}Z_{\beta}\left(\mu^{j}\left(N,x\right)\right)-\frac{\varkappa}{N}Z_{\beta}\left(2\pi-\mu^{N-1}\left(N,x\right)\right)-\varkappa Z_{\beta}\left(x\right)\right)
\]
 \[ =  \varkappa\sum_{j=0}^{N-2}\left(Z_{\beta}\left(\mu^{j}\left(N,x\right)\right)-Z_{\beta}\left(x\right)\right)-\underbrace{\varkappa Z_{\beta}\left(2\pi-\mu^{N-1}\left(N,x\right)\right)-\varkappa Z_{\beta}\left(x\right)}_{\equiv\mbox{I}}.
\]
Since part $\mbox{I}$, as well as its derivatives, is obviously uniformly
bounded in $N$ and $x,$ we restrict us to establish (\ref{eq:lower-bound-for-D2w})
for
 \[
\tilde{w}\left(N,x\right)=\sum_{j=0}^{N-2}\left[Z_{\beta}\left(\mu^{j}\left(N,x\right)\right)-Z_{\beta}\left(x\right)\right].
\]
We have
\[
\tilde{w}^{\prime}\left(N,x\right)  =  \sum_{j=0}^{N-2}\left[Z_{\beta}^{\prime}\left(\mu^{j}\left(N,x\right)\right)\left(\mu^{j}\left(N,x\right)\right)^{\prime}-Z_{\beta}^{\prime}\left(x\right)\right],
\]
\begin{eqnarray}
\tilde{w}^{\prime\prime}\left(N,x\right)  &=&  \sum_{j=0}^{N-2}\Bigl[
Z_{\beta}^{\prime\prime}\left(\mu^{j}\left(N,x\right)\right)\left(
\left(\mu^{j}\left(N,x\right)\right)^{\prime}\right)^{2} \label{eq:D2w-tilde}\\
&&
+Z_{\beta}^{\prime}\left(\mu^{j}\left(N,x\right)\right)\left(\mu^{j}\left(N,x\right)\right)^{\prime\prime}-Z_{\beta}^{\prime\prime}\left(x\right)\Bigr].\nonumber
\end{eqnarray}
To handle this, we need some uniformity-properties of $\mu^{j}\left(N,x\right).$
Elementary calculations give
\[
\left(\mu^{j}\left(N,x\right)\right)^{\prime}  =  \prod_{k=0}^{j-1}\mu^{\prime}\left(N,\mu^{k}\left(N,x\right)\right)=\prod_{k=0}^{j-1}\left(1+\frac{\varkappa}{N}Z_{\beta}^{\prime}\left(\mu^{k}\left(N,x\right)\right)\right),
\]
\[
\left(\mu^{j}\left(N,x\right)\right)^{\prime\prime} =  \sum_{l=0}^{j}\prod_{k=0,\, k\neq l}^{j-1}\left[\left(1+\frac{\varkappa}{N}Z_{\beta}^{\prime}\left(\mu^{k}\left(N,x\right)\right)\right)\right]
\]
\[\qquad \qquad \qquad \qquad
\times \frac{\varkappa}{N}Z_{\beta}^{\prime\prime}\left(\mu^{l}\left(N,x\right)\right)\left(\mu^{l}\left(N,x\right)\right)^{\prime}.\]
This implies that the following inequality
\begin{equation}
0<\left(\mu^{j}\left(N,x\right)\right)^{\prime}  <  \exp\left(\varkappa\zeta^{\prime}\right),
\quad 
\mbox{where}\quad
\zeta^{\prime}\equiv\max_{x\in\left[0,2\pi\right]}\left|Z_{\beta}^{\prime}\left(x\right)\right|
\label{eq:bounds-Dmu-j-01}
\end{equation}
holds for all large enough $N$. 
This again yields
\begin{eqnarray}
x\le\mu^{j}\left(N,x\right) & = & \mu^{j}\left(N,0\right)+\int_{0}^{x}\left(\mu^{j}\left(N,y\right)\right)^{\prime}dy\nonumber \\
 & \le & x+x \exp\left(\varkappa\zeta^{\prime}\right).\label{eq:bounds-for-mu-j-01}
\end{eqnarray}
Using this upper bound, we get some $N$-independent $\varepsilon_{0},$
such that for $x\in\left[0,\varepsilon_{0}\right]$ 
\[
Z_{\beta}^{\prime\prime}\left(\mu^{k}\left(N,x\right)\right)  >  0.
\]
This gives $N$-independent monotonicity of
\[
x\mapsto Z_{\beta}^{\prime}\left(\mu^{k}\left(N,x\right)\right)>0
\quad \mbox{for}\quad x\in\left(0,\varepsilon_{0}\right).
\]
Further, we can use (\ref{eq:bounds-for-mu-j-01}) to improve the
bounds (\ref{eq:bounds-Dmu-j-01}) for $\left(\mu^{j}\left(N,x\right)\right)^{\prime}$
in $x\in\left[0,\varepsilon_{0}\right]$ to 
\begin{equation}
1\le\left(\mu^{j}\left(N,x\right)\right)^{\prime}  <  \exp\left(\varkappa\zeta^{\prime}\right).\label{eq:bounds-Dmu-j-02}
\end{equation}
This implies \[
\mu^{j}\left(N,x\right)<x\cdot\exp\left(\varkappa\zeta^{\prime}\right).\]
We find
\[
0<\left(\mu^{j}\left(N,x\right)\right)^{\prime\prime}  \le  \frac{j\varkappa\zeta^{\prime\prime}}{N}\exp\left(2\varkappa\zeta^{\prime}\right)\le\varkappa\zeta^{\prime\prime}\exp\left(2\varkappa\zeta^{\prime}\right),\]
where \[
\zeta^{\prime\prime}\equiv\max_{x\in\left[0,2\pi\right]}\left|Z_{\beta}^{\prime\prime}\left(x\right)\right|.\]
Now observe that
\[
Z_{\beta}^{\prime\prime}\left(0\right)+Z_{\beta}^{\prime\prime\prime}\left(0\right)  =  \left(4-\frac{8}{\pi}\right)\beta^{2}+\frac{2}{\pi}\beta+\frac{1}{\pi}>0,
\]
i.e., eventually further decreasing of $\varepsilon_{0}>0$ gives,
with $\tilde{\varepsilon}_{0}=\varepsilon_{0}\cdot\exp\left(\varkappa\zeta^{\prime}\right):$
\begin{equation}
\min_{y\in\left[0,\tilde{\varepsilon}_{0}\right]}Z_{\beta}^{\prime\prime}\left(y\right)>-\min_{y\in\left[0,\tilde{\varepsilon}_{0}\right]}Z_{\beta}^{\prime\prime\prime}\left(y\right).\label{eq:D2Z-ge-D3Z}\end{equation}
Hence, for $x\in\left[0,\varepsilon_{0}\right]:$
\[
\tilde{w}^{\prime\prime}\left(N,x\right)  =  \sum_{j=0}^{N-2}\left(Z_{\beta}^{\prime\prime}\left(\mu^{j}\right)
\left(\left(\mu^{j}\right)^{\prime}\right)^{2}+
\underbrace{Z_{\beta}^{\prime}\left(\mu^{j}\right)
\left(\mu^{j}\right)^{\prime\prime}}_{\ge0}-Z_{\beta}^{\prime\prime}\left(x\right)\right)
\]
\[
 \quad \ge  \sum_{j=0}^{N-2}\left(Z_{\beta}^{\prime\prime}
\left(\mu^{j}\right)\left(\left(\mu^{j}\right)^{\prime}\right)^{2}-Z_{\beta}^{\prime\prime}\left(\mu^{j}\right)+
\int_{x}^{\mu^{j}}Z_{\beta}^{\prime\prime\prime}\left(y\right)dy\right)
\]
\[ 
\quad =  \sum_{j=0}^{N-2}\left(Z_{\beta}^{\prime\prime}\left(\mu^{j}\right)
\left(\left(\left(\mu^{j}\right)^{\prime}\right)^{2}-1\right)+
\int_{x}^{\mu^{j}}Z_{\beta}^{\prime\prime\prime}\left(y\right)dy\right)
\]
 \[ 
\quad \ge  \sum_{j=0}^{N-2}\left(\min_{y\in\left[0,\tilde{\varepsilon}_{0}\right]}Z_{\beta}^{\prime\prime}\left(y\right)
\left(\left(\left(\mu^{j}\right)^{\prime}\right)^{2}-1\right)+\min_{y\in\left[0,\tilde{\varepsilon}_{0}\right]}Z_{\beta}^{\prime\prime\prime}
\left(y\right)\left(\mu^{j}-x\right)\right),
\]
where we have omitted the arguments $(N,x)$ of $\mu$ for brevity.
Using (\ref{eq:D2Z-ge-D3Z}), we continue the estimations
\[
\cdots  \ge  \sum_{j=0}^{N-2}\left(\min_{y\in\left[0,\tilde{\varepsilon}_{0}\right]}Z_{\beta}^{\prime\prime}\left(y\right)
\left(\left(\left(\mu^{j}\right)^{\prime}\right)^{2}-1-\left(\mu^{j}-x\right)\right)\right)
\]
\[
\quad  =  \min_{y\in\left[0,\tilde{\varepsilon}_{0}\right]}Z_{\beta}^{\prime\prime}\left(y\right)
\sum_{j=0}^{N-2}\left(\left(\underbrace{\left(\mu^{j}\right)^{\prime}}_{\ge1}
\right)^{2}-1-\int_{0}^{x}\left(\underbrace{\left(\mu^{j}
\right)^{\prime}}_{\le\mu^{j}}-1\right)dy\right)
\]
\[
 \quad \ge  \min_{y\in\left[0,\tilde{\varepsilon}_{0}\right]}Z_{\beta}^{\prime\prime}
\left(y\right)\sum_{j=0}^{N-2}\left(\left(\left(\mu^{j}\right)^{\prime}\right)-1-x\left(\left(\mu^{j}\right)^{\prime}-1\right)\right)
\]
\[
\quad \ge  \min_{y\in\left[0,\tilde{\varepsilon}_{0}\right]}Z_{\beta}^{\prime\prime}\left(y\right)
\sum_{j=0}^{N-2}\left(1-x\right)\left(\left(\mu^{j}\right)^{\prime}-1\right) \ge 0.
\]
This establishes (\ref{eq:lower-bound-for-D2w}) and hence $Y_{1}\left(N,x\right)>x\in\left[0,\varepsilon_{0}\right]$
for large enough $N.$
$\square$

 \bibliographystyle{wivchnum}

\end{document}